\documentstyle[12pt]{article}

\renewcommand{\baselinestretch}{1.2}

\begin{document}

\renewcommand{\baselinestretch}{1.07}

\title{ \Large \bf Cosmic cristallography:\\
                   three multi-purpose functions\\
                         }
\author{A. Bernui\thanks{Permanent address:
                          Facultad de Ciencias, Universidad Nacional de Ingenier\'{\i}a, 
                          Apartado 31--139, Lima 31 -- Peru.
                          {\sc e-mail:}  bernui@fc-uni.edu.pe }  \\
                  and  \  
         A.F.F. Teixeira\thanks{{\sc e-mail:} teixeira@cbpf.br} \\ 
         \\  
         Centro Brasileiro de Pesquisas F\'\i sicas\\
         Departamento de Relatividade e Part\'\i culas \\
         Rua Dr.\ Xavier Sigaud 150 \\
         22290-180 Rio de Janeiro -- RJ, Brazil \\ \\
       }

\date{\today}  

\maketitle

\begin{abstract}
A solid sphere is considered, with a uniformly distributed infinity 
of points. Two points being pseudorandomly chosen, the analytical 
probability density that their separation have a given value is computed, 
for three types of the underlying geometry: $E^3, H^3$ and $S^3$. 
Figures, graphs and histograms to complement this short note are given. 
\end{abstract}
\newpage
\section{Introduction} 
\setcounter{equation}{0}

The shape of the universe is presently under investigation,
and cosmic crystallography (CC) is one of the various methods
proposed to determine it \cite{LeLaLu96}.

The idea which supports the CC method is that if the universe is
multiply connected then multiple images of a same cosmic object
(a given quasar, say) may be observed in the sky.
The separations between pairs of these images are correlated by
the geometry and the topology of the spacetime;
so if one selects a catalogue of observed images of cosmic objects
and performs a histogram of the separations $l$ between the images,
then these correlations manifest either as spikes (associated 
with Clifford translations) or as slight deformations of the histogram
of the corresponding simply connected manifold \cite{spikes99}.

A variant method was recently proposed by Fagundes and Gausmann
~\cite{FagundesGausmann98} , of subtracting from a histogram
$ \phi (l_{i}) $
of a multiply connected space a histogram of the corresponding
simply connected space. The reported result was a plot with much
oscillations in small scales.

In the present paper we propose an alternative to Fagundes-Gausmann
method: we derive continuous probability density functions
$ {\cal F}(l) $        to be subtracted from the histogram
$ \phi (l_{i}) $ ,        and thereby obtain a histogram
sensibly more suitable for analysis.

In section 2 we derive the functions $  {\cal F}(a,l) $
for the Euclidean, hyperbolic and elliptic geometries,
and in section 3 we make a few comments.

\section{Probability densities}
\setcounter{equation}{0}

We consider one of the simply connected spaces
$ E^3, H^3 $ or $ S^3 $. In that space, a spherical solid ball
$ {\cal B}_a $ is taken , with radius $a$.
The ball is assumed to contain an infinite number of pointlike objects,
spatially distributed as uniformly as possible.
We next select pseudorandomly two points of 
$ {\cal B}_{a} $ and ask for the probability 
$ {\cal F}(a,l)dl $ that the separation between the points
lie between $l$ and $l + dl$.

\vspace{5mm}

{\bf a. Euclidean geometry}

In the Euclidean three-space take a ball $ {\cal B}_{a} $
centred at the origin $ O $ and 
select two points $ P , Q $ in the ball; let
$ r\in [0 ,a] $ be the radial position of $P$ and let
$ l\leq 2a $ be the distance from $P$ to $Q$ (see Figure 1).

Clearly the probability density 
$ {\cal F}_E(a,r,l)$ of this configuration is proportional
both to the area $ {\cal S}_E(r)= 4\pi r^2 $ of the locus of $P$ well as 
to the area of the locus of $Q$. 
When $r + l < a$ this locus is a sphere $S^2$ with area ${\cal S}_E(l)$, 
while when $r+l>a$ the locus is a spherical disk $D^2$ in $E^3$ with area 
$D_E(a,r,l) = (\pi l/r)[a^2-(l-r)^2]$. 

The probability density of the configuration is then 
\begin{equation}
\label{euc1}
{\cal F}_E(a,r,l)=k{\cal S}_E(r)\{{\cal S}_E(l)\times\Theta (a-l-r)+D_E(a,r,l)
\times\Theta (l+r-a)\}, 
\end{equation}
where $k$ is a normalization constant and  $\Theta$ is the Heaviside function. 

We integrate eq. (\ref{euc1}) for $r\in [0,a]$, and finally obtain, 
for $l\in (0, 2a]$,
\begin{equation}
\label{euc2}
{\cal F}_E(a,l)=\frac{3 l^2}{16 a^6} (2a-l)^2 (l+4a),
\end{equation} 
where the value $k=9/(16\pi ^2 a^6)$ was set to satisfy the normalization condition 
\begin{equation}
\label{euc3}
\int_{0}^{2a} {\cal F}_E(a,l)dl = 1. 
\end{equation}

One often encounters in the literature reference to the probability density 
${\cal P}(s)$ that the {\em squared} separation be $s$; since $s=l^2$ and 
${\cal P}(s)ds = {\cal F}(l)dl$, then 
\begin{equation}
\label{euc4}
{\cal P}_E(a,s) = \frac{3\sqrt{s}}{32 a^6}(2a-\sqrt{s})^2 (\sqrt{s}+4a),
\end{equation} 
valid for $s\in (0, 4a^2] $. 

In Figure 2 we reproduce a typical mean pair separation histogram (MPSH) for 
pseudorandomly distributed objects in an Euclidean solid sphere ${\cal B}_a$ 
with arbitrary radius, 
together with the corresponding probability density ${\cal F}_E(a,l)$. 
It should be stressed that, differently from the hyperbolic and elliptical 
cases, the shape of the function ${\cal F}_E(a,l)$ does not depend on 
the value of the radius $a$.

\vspace{5mm}

{\bf b. Hyperbolic geometry}

To obtain the probability density ${\cal F}_H(a,l)$ for the hyperbolic geometry 
we follow the same lines as before. The area of a sphere with radius $r$ is now 
$S_H(r)=4\pi R^2 \sinh^2r/R$, where $R$ is the radius of curvature of the 
geometry; without loss of generality we henceforth set $R=1$. 
On the other hand, the area of a spherical disk $D^2$ in $H^3$ is (see Figure 1) 
\begin{equation}
\label{hip1} 
D_H(a,r,l)= 2\pi \sinh\! l [\:\sinh\! l -\cosh\! l  \coth\! r + \cosh\! a \:
{\rm csch} r ],
\end{equation} 
to be considered whenever $r+l>a$. 

The probability density of the configuration is then 
\begin{equation}
\label{hip2}
{\cal F}_H(a,r,l)=kS_H(r)[S_H(l)\times \Theta(a-l-r) + D_H(a,r,l)\times 
\Theta(l+r-a)], 
\end{equation}
which upon integration for $r\in [0,a]$ and normalization gives 
\begin{equation}
\label{hip3}
{\cal F}_H(a,l)=\frac{8 \sinh^2 l}{(\sinh 2a -2a)^2}[\cosh\! a \:{\rm sech}(l/2)\sinh(a-l/2)-(a-l/2)], 
\end{equation}
valid for $l\in (0,2a]$. 

In Figure 3 we reproduce graphs of ${\cal F}_H(a,l)$ for three values of the radius $a$. 
For $a<\!<1$ the function tends to the Euclidean one given in Figure 2, as expected. 
For increasing values of $a$ the function shifts towards the large values of $l$, 
and for $a>\!>1$ a strong concentration of ${\cal F}_H(a,l)$ is found near the 
extreme value $l=2a$. 

Figure 4 shows a typical MPSH in the hyperbolic three-space, together with 
the corresponding probability density ${\cal F}_H(a,l)$.

\vspace{5mm}

{\bf c. Elliptic geometry} 

The basic strategy to obtain the probability density ${\cal F}_S(a,l)$ 
is the same as before, and the calculations are similar whenever the diameter $2a$ 
of the ball is less than the separation $\pi R$ between antipodal points 
in the three-sphere; we then find for  ${\cal F}_S(a,l)$ the expression, 
valid for $l\in (0,2a]$, 
\begin{equation}
\label{eli1}
{\cal F}_S(a<\pi/2,l)=\frac{8 \sin^2 l}{(2a-\sin 2a)^2}[(a-l/2)-\cos \!a\:\sec (l/2)\:\sin(a-l/2)], 
\end{equation} 
where we have taken $R=1$ without loss of generality.

However, the cases where $a>\pi /2$ are considerably trickier to deal with, 
due to the treacherous connectivity of the spherical space $S^3$ 
and the new requirement that $l$ must not exceed $\pi$. 
A much larger quantity of trivial integrations now comes into scene, 
before the following expression is eventually obtained: 
\begin{eqnarray}
{\cal F}_S(a,l) & = & \frac{8\sin^2 l}{[2a-\sin 2a]^2}\{[2a-\sin 2a-\pi]
                 + \label{eli2} \\  
       &   & \Theta(2\pi-2a-l)\times [\:\sin\! 2a+\pi-a-l/2-\cos\! 
           a\: \sec(l/2)\:\sin(a-l/2)]\}\nonumber,
\end{eqnarray}
valid for all $a\in (0,\pi]$ and $l\in (0,{\rm min}(2a,\pi)]$.  

In Figure 5 four graphs of ${\cal F}_S(a,l)$ are shown, for different values 
of the radius $a$ of the ball. 
For $a$ increasing from $0$ to $\pi$ the function continuously shifts towards 
the smaller values of $l$. 
In particular, when $a=\pi/2$ we have 
\begin{equation}
\label{eli3}
{\cal F}_S(\pi/2,l)=\frac{4}{\pi}(1-l/\pi)\sin^2\! l\; , 
\end{equation} 
while when $a=\pi$ we have the harmonic, symmetric probability density  
\begin{equation}
\label{eli4}
{\cal F}_S(\pi,l)=\frac{2}{\pi}\sin^2\! l\; .
\end{equation}

In Figure 6 a sample MPSH in the spherical space is reproduced, together with the corresponding probability density ${\cal F}_S(a,l)$.

\section{Comments} 
It is perhaps worth clarifying the meaning of the functions ${\cal F}(a,r,l)$: 
if we pseudorandomly choose two points $P$ and $Q$ in a solid sphere with radius $a$ 
(see Figure 1) then ${\cal F}(a,r,l)\:dr\:dl$ is the probability that $P$ lies 
between the radial positions $r$ and $r+dr$, times the probability that 
the separation from $Q$ to $P$  lies between the values $l$ and $l+dl$. 
The form of ${\cal F}(a,r,l)$ clearly depends on the geometry one is concerned with. 

Each histogram in Figures 2, 4, and 6 has $m=100$ subintervals and is a MPSH -- 
a mean pair separation histogram comprising $K=10$ comparable catalogues with $N=50$ 
objects each \cite{spikes99}. 
All computer-generated histograms assume a homogeneous distribution, 
as described by Lehoucq, Luminet and Uzan \cite{LeLuUz98}. 

A close inspection of Figure 2 shows that the most probable separation 
between two arbitrarily chosen points in an Euclidean solid ball is slightly 
greater than the radius of the ball; also the maximum of ${\cal P}_E(a,s)$ 
in eq.(\ref{euc4}) occurs when   
$s/4a^2=0.134$, 
in agreement with plots of Fagundes and Gausmann \cite{FagundesGausmann98b}. 

A characteristic feature of the hyperbolic geometries is that at large distances 
there is more space than in the Euclidean geometries;
this fact is clearly exhibited in Figure 3, which shows a strong predominance of 
large separations $l$ when the radius $a$ of the solid sphere is large. 
It is worth noting that Fagundes and Gausmann \cite{FagundesGausmann98} 
obtained histograms with $a=2.34$ 
which closely resemble ours with $a=2$ in Figure 4. 
In contrast, the hyperbolic histograms given by Lehoucq {\em et al.} \cite{LeLuUz98} 
use radius $a$ nearly $0.6$, so they are similar to our Euclidean one.  

Oppositely to the hyperbolic case, in distant places endowed with the {\em elliptic} geometry there is {\em less} space than in the Euclidean geometry; 
this is nicely illustrated in Figure 5. 
Indeed, with increasing $a$ ( increasing solid ball) the probability of finding 
small distances $l$ (in comparison with $a$) also increases. 
We further note that when $a$ increases from $\pi/2$ to $\pi$ the ratio 
$l_{max}/2a$ recedes from $1$ to $1/2$.

\newpage
{\large {\bf Captions for the Figures}} 

\vspace{5mm}

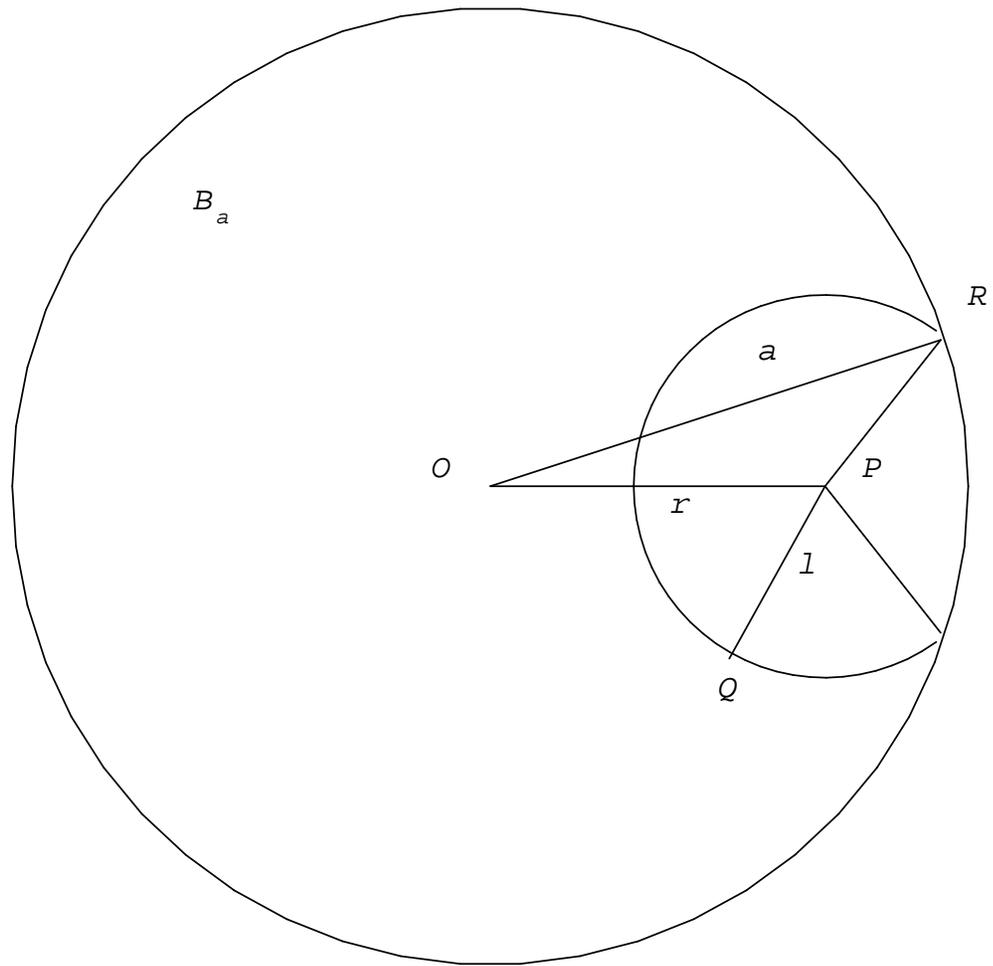
Figure 1. A two-dimensional picture of the solid ball ${\cal B}_a$. We have $OR=a$ 
(the radius of ${\cal B}_a$), $OP=r$, $PQ=l$. The circular arc with centre $P$ 
represents a spherical disk $D^2$; the disk becomes a sphere $S^2$ whenever 
$r+l\leq a$.

\vspace{5mm}
Figure 2. In the Euclidean space $E^3$, a sample MPSH for a solid ball with arbitrary 
radius $a$, together with the corresponding probability density 
eq.~(\ref{euc2}).

\vspace{5mm}
Figure 3. In the hyperbolic space $H^3$ with curvature radius $R=1$, the probability 
densities eq.(\ref{hip3}) for solid spheres with radii $a=0.01$ (dots), 2.0 (line) 
and 10 (bold line).       

\vspace{5mm}
Figure 4. A sample MPSH and the corresponding probability density eq.(\ref{hip3}) 
for a solid sphere with radius $a=2$ in $H^3$ with unitary radius.

\vspace{5mm}   

Figure 5. In the three-sphere $S^3$ with radius $R=1$, the probability densities 
eqs.(\ref{eli2})--(\ref{eli4}) for balls with radii $a=\pi$ (bold line),    
$3\pi/4$, $\pi/2$, and $0.01$ (dots).

\vspace{5mm}
Figure 6. In the spherical space $S^3$ with unitary radius, a sample MPSH 
for a solid ball with radius $a=\pi/2$, and the corresponding probability 
density eq.(\ref{eli3}).

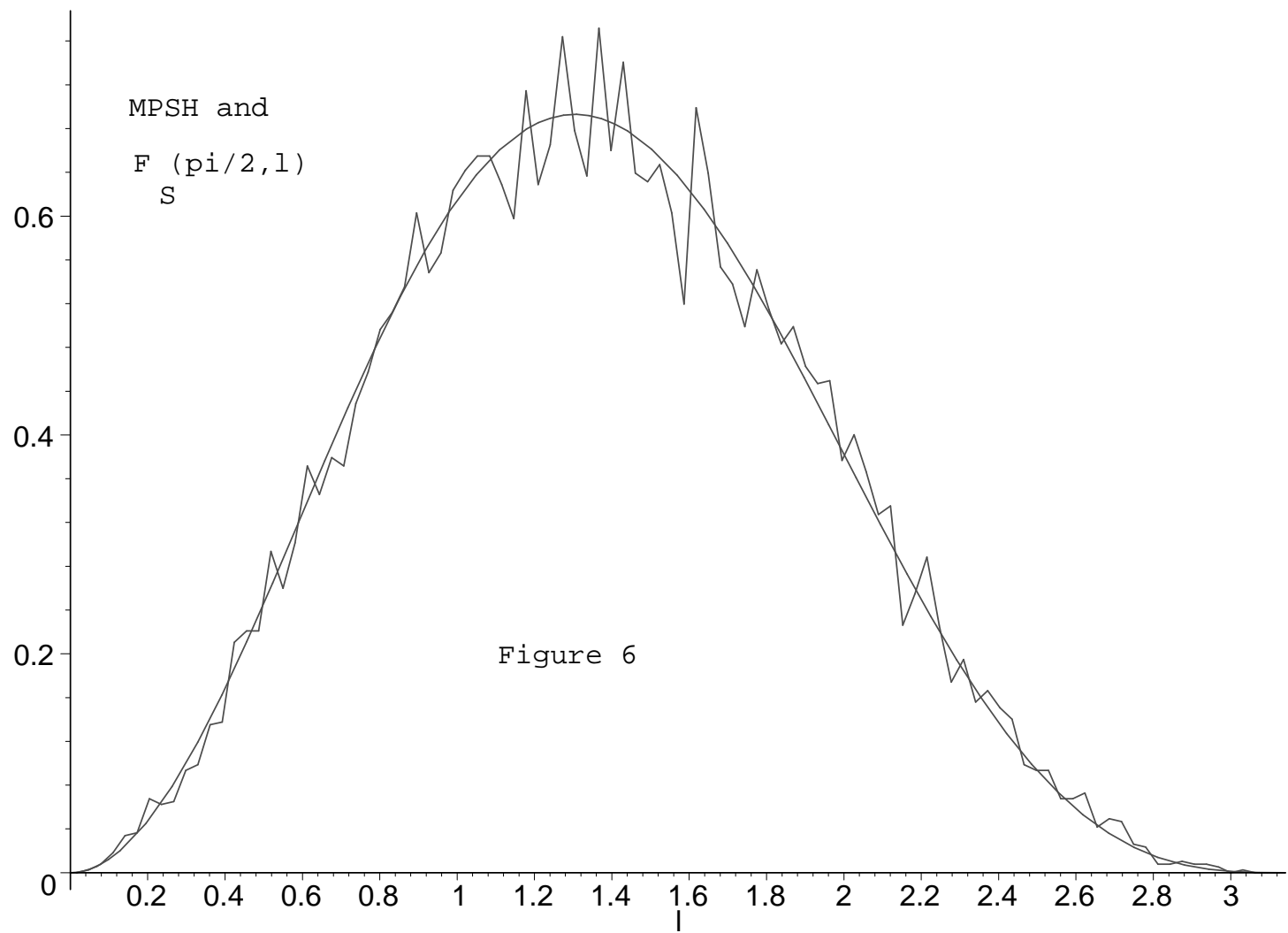
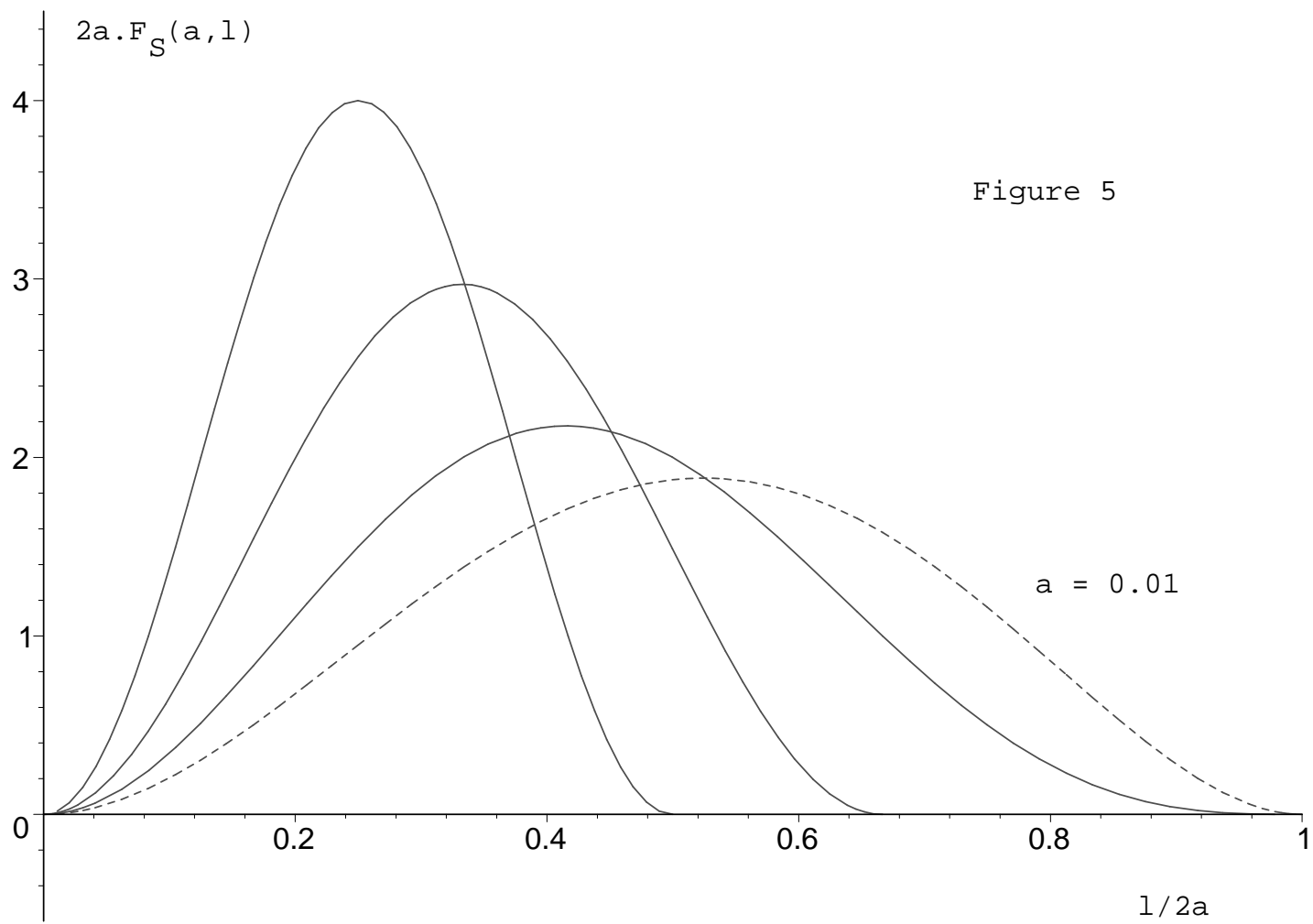
\end{document}